\documentclass[aps,pra,twocolumn,groupedaddress]{revtex4-2}
\usepackage{graphicx}
\usepackage{caption}
\DeclareCaptionLabelSeparator{dot}{. }
\makeatletter
\def\justified{
	\let\\\@normalcr
	\@rightskip\z@skip \rightskip\@rightskip
	\leftskip\z@skip
	\parindent 0em\relax
	\setlength{\parfillskip}{0pt plus 1fil}}
\DeclareCaptionJustification{justified}{\justified}
\usepackage{subcaption}
\usepackage{amsmath}
\usepackage{amssymb}
\usepackage{mathrsfs} 
\usepackage{braket}
\usepackage{ragged2e}
\usepackage[colorlinks,urlcolor=blue ,citecolor=blue ,linkcolor=blue ]{hyperref}
\usepackage{xcolor}
\usepackage{nicefrac} 
\usepackage{lipsum}
\usepackage{nameref}
\usepackage{hyperref}
\usepackage{textcomp} 
\usepackage{urwchancal}
\usepackage{xcolor}
\usepackage{float}
\definecolor{darkgreen}{rgb}{0,0.5,0}
\usepackage{multirow}
\usepackage{tabularx}
\usepackage[normalem]{ulem}
\usepackage{siunitx}

\newcommand{\um}{\mu{\rm m}}

\newcommand{\as}{\ensuremath{a_{s}}}

\newcommand{\edd}{\ensuremath{\epsilon_{\rm dd}}}

\newcommand{\Dyb}{\ensuremath{^{164}}{\rm Dy}}
\newcommand{\Dybs}{\ensuremath{^{164}}{\rm Dy }}

\usepackage[separate-uncertainty=true, list-units=repeat,multi-part-units=brackets,exponent-product=\cdot]{siunitx}

\begin{document}

\title{Characterizing supersolid phase transitions in trapped quantum gases with tilted dipoles at finite temperature
} 

\author{Julian Kusch$^{1}$}
\author{Lauriane Chomaz$^{1}$}\email[]{chomaz@uni-heidelberg.de}
\author{Wyatt Kirkby$^{1}$}\email[]{wyatt.kirkby@gmail.com}

	\affiliation{$^{1}$ Physikalisches Institut, Universität Heidelberg, Im Neuenheimer Feld 226, 69120, Heidelberg, Germany.}	

	\date{\today}
	
\begin{abstract}
Motivated by a recent experimental work~[\href{https://arxiv.org/abs/2608.20327}{arXiv:2608.20327}], we theoretically investigate the nature of phase transitions in a trapped dipolar quantum gas upon tuning the contact interaction strength and dipole orientation. Using extended Gross-Pitaevskii simulations including thermal and quantum effects, we map the phase diagram comprising unmodulated superfluids, triangular droplet arrays, and stripe states. We further characterize the order of the transitions separating them through imaginary-time hysteresis analysis. The unmodulated-to-modulated transition evolves from discontinuous to continuous with increasing dipole tilt before becoming discontinuous again at large tilts. The degree of continuity is further affected by particle number, temperature, and confinement. While continuous behavior generally correlates with the emergence of stripe states, this correspondence is not systematic, and finite-size effects also influence the continuity of the transition. The triangular-to-stripe transition is found to be discontinuous at small contact interaction strengths and becomes increasingly continuous upon approaching the unmodulated-to-modulated phase boundary. Our results reveal a tunable transition order in finite dipolar quantum gases and provide guidance for its experimental investigation.
\end{abstract}

\maketitle

\section{Introduction}\label{sec:intro}

The theory of phase transitions describes how distinct macroscopic phases emerge as external parameters are varied~\cite{Goldenfeld2018lop}. 
Within the Landau paradigm, a phase transition is characterized by an order parameter, and, in the thermodynamic limit, it is associated with a nonanalyticity of the free energy. 
The nature of this nonanalyticity determines the order of the transition and governs the system's behavior in its vicinity. While more general types of phase transitions exist, conventional symmetry-breaking transitions are commonly classified as first- or second-order. 
A first-order transition is characterized by a discontinuity in the first derivative of the free energy. 
This typically entails a discontinuous change of the order parameter at the transition, the metastability of the competing phases in the vicinity of the transition, and hysteresis upon its dynamical crossing. In contrast, second-order transitions exhibit higher-order nonanalyticities of the free energy, are characterized by a continuous evolution of the order parameter, and are associated with critical scaling behaviors defining broad universality classes. 

Phase transitions become notably rich when several orders compete within the same system. 
Supersolids provide an ideal setting in this respect, combining superfluidity and crystalline orders with potentially different spatial arrangements~\cite{Gross1957Unified,Gross1958Classical,Boninsegni2012Colloquium}.
Such a rich landscape is especially well realized and controlled in dipolar quantum gases~\cite{Chomaz2022Dipolar,Chomaz2026Quantum,Chomaz2019LongLived,Tanzi2019Observation,Boettcher2019Transient,Sohmen2021Birth, Norcia2021Twodimensional,Biagioni2022Dimensional,Bland2022TwoDimensional,He2025Observation,Chandrashekara2026Competing}. 
Supersolid states have been realized in magnetic-gas experiments in the form of one- and two-dimensional droplet arrangements with dipoles polarized along the tight confinement axis~\cite{Norcia2021Twodimensional,Bland2022TwoDimensional,Tanzi2019Observation,Boettcher2019Transient,Chomaz2019LongLived,Chomaz2022Dipolar,Chomaz2026Quantum,Biagioni2022Dimensional} as well as, most recently, two-dimensional  droplet and stripe states with tilted dipoles~\cite{He2025Observation, Chandrashekara2026Competing}.
In the latter case, transitions between unmodulated superfluids, triangular supersolids, stripe supersolids, and the corresponding insulating states were explored in a single system upon tuning dipole tilt and interaction strength~\cite{Chandrashekara2026Competing}.

These experimental advances have been accompanied by extensive theoretical efforts to understand the stability, structure, and phase diagrams of dipolar supersolids.
Their description largely relies on the extended Gross-Pitaevskii equation (eGPE), an effective mean-field approach incorporating beyond-mean-field corrections from quantum fluctuations, which are essential for stabilizing droplets and supersolid states~\cite{Waechtler2016Ground,Waechtler2016Quantum,Baillie2016Self,Schmitt2016Self,Chomaz2016QuantumFluctuationDriven}.  
This framework has been used extensively to map the zero-temperature phase diagrams of dipolar gases~\cite{ Baillie2018Droplet, Roccuzzo2019Supersolid, Zhang2019Supersolidity,Poli2021Maintaining, Biagioni2022Dimensional, Ripley2023Twodimensional, Zhang2024Metastable,Lima2025Supersolid,SanchezBaena2025Tilted}.
More recently, thermal fluctuations have also been included in finite-temperature extensions of the eGPE~\cite{SanchezBaena2023Heating, He2025Accessing} and their impact on the phase diagram has been studied~\cite{Bombin2025Creating, Sanchez2026Quasi, Yu2025Thermally, Sohmen2021Birth, SanchezBaena2026FiniteTemperature}. Depending on contact and dipolar interaction strengths, dipole orientation, density, confinement, and temperature, the gas can evolve between unmodulated superfluid and density-modulated states with a rich variety of crystalline structures, varying from triangular to stripe and labyrinthine states, or even honeycomb for increasing densities~\cite{Lu2015,Zhang2019Supersolidity,Norcia2021Twodimensional,Hertkorn2021Pattern,Bland2022TwoDimensional,Ripley2023Twodimensional,He2025Accessing}.

The nature of the phase transitions connecting these different states has also attracted significant theoretical interest. 
Studies in infinite systems have shown that the order of the unmodulated-to-modulated transition can vary depending on the gas parameters and geometry. 
For one-dimensional crystals, it can be continuous or discontinuous depending on the gas density~\cite{Blakie2020Supersolidity}. 
For two-dimensional crystalline structures, triangular arrangements generally favor a first-order transition, whereas stripes, which are effectively one-dimensional modulations, may yield continuous transitions~\cite{Lu2015,Zhang2019Supersolidity,Ripley2023Twodimensional,SanchezBaena2024SuperfluidSupersolid,Biagioni2022Dimensional,Lima2025Supersolid}. 
Further studies have investigated the influence of finite size and trapping on the continuity of the transition~\cite{Biagioni2022Dimensional, Poli2021Maintaining,Shi2026MatterWave}, and of finite temperature on the phase boundaries
~\cite{Sanchez2026Quasi, Yu2025Thermally}. The triangular-to-stripe transition has also been investigated in the thermodynamic limit~\cite{Zhang2019Supersolidity,Ripley2023Twodimensional,Lima2025Supersolid}. It is generally first-order but has been predicted to become second-order with increasing dipole tilt, depending on the density~\cite{Lima2025Supersolid}. 
However, the combined influence of dipole tilt, temperature, particle number, and confinement on the character of the different transitions remains largely unexplored.

In this work, motivated by the experimental exploration of Ref.~\cite{Chandrashekara2026Competing}, we investigate how the nature of the phase transitions in a trapped dipolar gas evolves with dipole tilt and contact interaction strength, for varying trap parameters, temperature, and atom number. In particular, we consider parameter ranges consistent with the experimental values of Ref.~\cite{Chandrashekara2026Competing}.
In Sec.~\ref{sec:system}, we introduce the system and the extended Gross-Pitaevskii framework used to model it, including quantum and finite-temperature corrections. 
In Sec.~\ref{sec:phasediagram}, we explore the phase diagram of our trapped gas of magnetic atoms. We investigate the unmodulated-to-modulated and triangular-to-stripe transitions via hysteresis between opposite imaginary-time propagation directions in Sec.~\ref{subsec:unmod_transition} and~\ref{subsec:stripe_transition}, respectively. 
Finally, in Sec.~\ref{subsec:large_tilts}, we discuss the distinct unmodulated-to-modulated transition behavior found at large tilt angles. Sec.~\ref{sec:conclusion} summarizes the results and gives directions for further studies.  

\section{System and Model}\label{sec:system}

\begin{figure}[t!]
    \centering
    \includegraphics[width=0.5\linewidth]{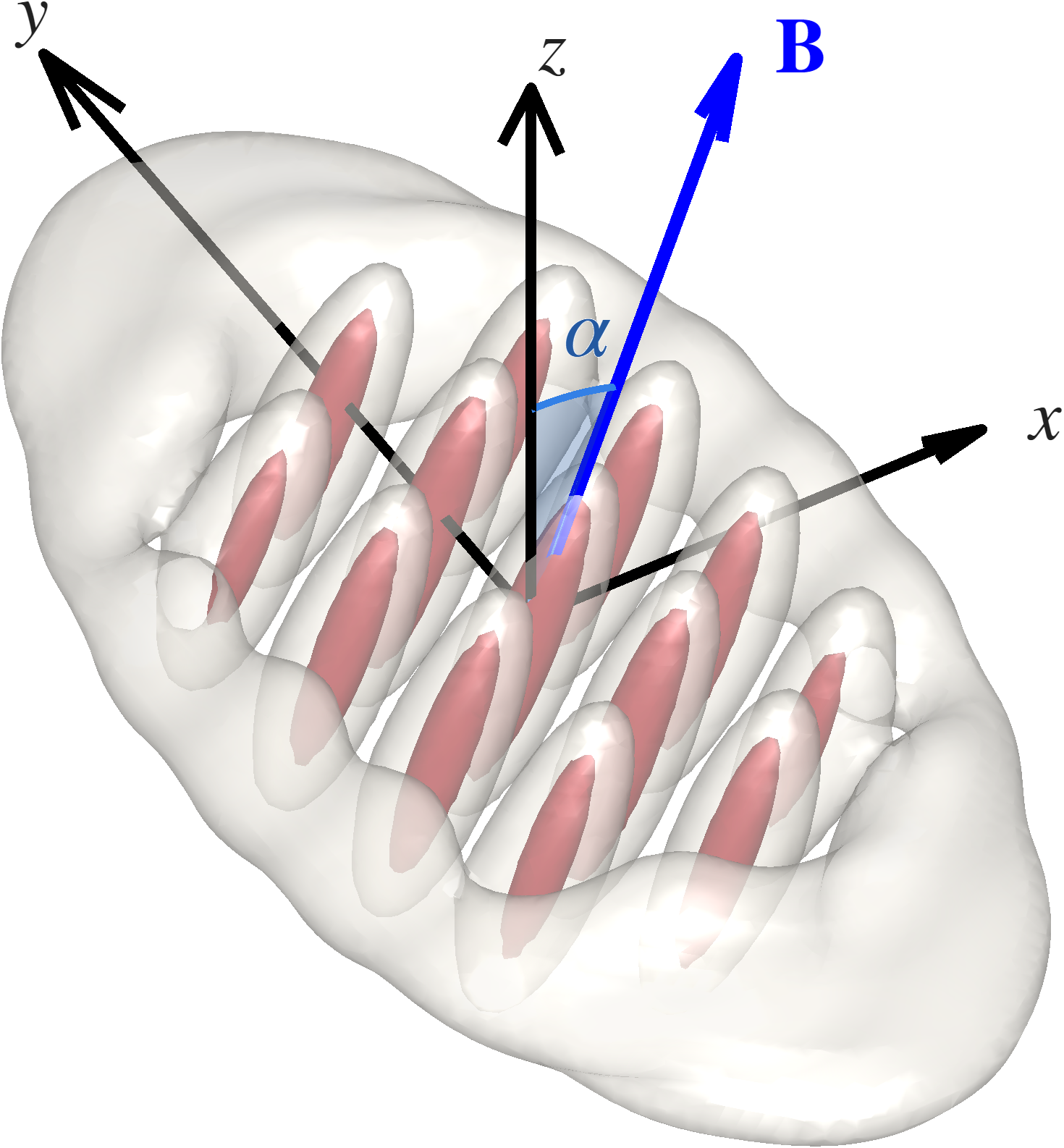}
    \caption{\textbf{System Geometry.} Sketch of the dipolar quantum gas in a surfboard-shaped trap. The shaded areas show isosurfaces of the density of a modulated ground state obtained from the TeGPE. The harmonic trap axes are aligned along the $x,y,z$ coordinate axes. The polarization axis is given by the tilt $\alpha$ of the magnetic field in the $x-z$ plane relative to the tightest confinement axis $z$.}
    \label{fig:geometry}
\end{figure}

We consider a dipolar quantum gas of \Dybs atoms, which we model using a finite-temperature extended Gross-Pitaevskii equation (TeGPE),
\begin{equation}
    \mathrm{i}\hbar\frac{\partial \Psi(\textbf{r},t)}{\partial t}=\Biggl[H_0+H_{\mathrm{int}}+H_{\mathrm{Q}}+H_{\mathrm{T}}\Biggr]\Psi(\textbf{r},t),\label{eq:TeGPE}
\end{equation}
with the mean-field wave function $\Psi$, normalized to the particle number $N$, 
\begin{align}
    \int \mathrm{d}\textbf{r}\left|\Psi(\textbf{r})\right|^2 = N.
\end{align}
The terms
\begin{equation}
    H_0=\;-\frac{\hbar^2\nabla^2}{2m}+V(\textbf{r})\;,\label{eq:H0}
\end{equation}
    and
\begin{equation}
    H_{\mathrm{int}}=\;\int\mathrm{d}\textbf{r}'U(\textbf{r}-\textbf{r}')|\Psi(\textbf{r}',t)|^2\label{eq:Hint}
\end{equation}
are the single-particle and two-body terms, respectively. 
The external trapping potential, $V(\textbf{r})=\frac{1}{2}m(\omega_x^2x^2+\omega_y^2y^2+\omega_z^2z^2)$, is a surfboard-shaped harmonic potential with fixed planar frequencies $(\omega_x, \omega_y)=2\pi\times(\nu_x, \nu_y) = 2\pi\times(40, 20)\,\mathrm{Hz}$. 
The strongest confinement is given by the vertical trapping frequency, $\omega_z=2\pi\times\nu_z$, where $\nu_z$ is varied in the paper in the range $[100, 200]\,\mathrm{Hz}$. 
A reference value is chosen with $\nu_z= \SI{150}{\Hz}$. 
The two-body interaction potential reads
\begin{equation}
    U(\textbf{r}) = g\delta(\textbf{r})+\mu_0\mu_m^2\frac{1-3\cos^2\vartheta}{4\pi|\textbf{r}|^3}\label{eq:Uint}\;,
\end{equation}
where $g=4\pi\hbar^2a_s/m$ characterizes the strength of the contact interaction with $a_s$ being the $s$-wave scattering length, $m$ is the atomic mass, $\mu_0$ is the vacuum magnetic permeability, $\mu_m$ the magnitude of the magnetic dipole moment, and $\vartheta$ is the angle of $\textbf{r}$ with respect to the polarization axis.
The strength of the dipolar interactions can be characterized by the dipolar length, $a_\mathrm{dd} =\mu_0\mu_m^2m/(12\pi\hbar^2)$. 
For \Dyb, $a_\mathrm{dd}= 131\,a_0$, where $a_0$ is the Bohr radius.
Throughout the manuscript, we tune the dipole tilt angle $\alpha$ relative to $z$-axis (see Fig.~\ref{fig:geometry}), which can be experimentally achieved by tuning the magnetic field $\textbf{B}$ with respect to the fixed trap orientation~\cite{Chandrashekara2026Competing,Wenzel2017Striped}.

The beyond-mean-field quantum fluctuation contribution,
\begin{equation}
    H_{\mathrm{Q}}=\gamma_{\mathrm{Q}}|\Psi(\textbf{r},t)|^3\label{eq:HLHY}\;,
\end{equation}
helps stabilize supersolid and droplet states beyond the mean-field instability \cite{Kadau2016Observing,FerrierBarbut2016Observation,Schmitt2016Self,Chomaz2016QuantumFluctuationDriven,Waechtler2016Quantum,Waechtler2016Ground,Bisset2016Groundstate}. Here, $\gamma_{\mathrm{Q}}=\frac{32}{3}g\sqrt{a_s^3/\pi}(1+3\edd^2/2)$ \cite{Schuetzhold2006Meanfield,Lima2011Quantum,Lima2012Beyond}, where $\edd=a_\mathrm{dd}/a_s$ captures the relative strength of the long-range dipolar interaction to the short-range contact interaction.
This treatment of the quantum fluctuations relies on a local-density approximation of the leading-order correction to the mean-field energy density due to condensate depletion.
For $\edd>1$, this contribution becomes complex, but its imaginary component is typically neglected as it remains small for the interaction strengths and densities relevant for atomic systems. 
The quadratic form of $\gamma_{\mathrm{Q}}$ given above corresponds to an approximation to the exact hypergeometric integral relevant for $\edd\sim1$~\cite{FerrierBarbut2016Observation,Bisset2016Groundstate,Ripley2023Twodimensional}.

The effect of thermal fluctuations on the ground state energy, captured by $H_{\mathrm{T}}$, is calculated by considering finite-temperature contributions to the Bogoliubov quasiparticle occupation number in the grand-canonical ensemble~\cite{Pitaevskii16} and by using a local-density approximation~\cite{SanchezBaena2023Heating,SanchezBaena2024SuperfluidSupersolid,He2025Accessing},
\begin{equation}
    H_{\mathrm{T}}=\int\frac{\mathrm{d}\textbf{k}}{(2\pi)^3}\frac{\tilde{U}(\textbf{k})}{(\mathrm{e}^{\beta \epsilon_\textbf{k}}-1)}\frac{\tau_\textbf{k}}{\epsilon_\textbf{k}}\label{eq:HTherm}\;,
\end{equation}
where the free-particle dispersion is $\tau_\textbf{k}=\hbar^2k^2/(2m)$ and $\beta = 1/(k_\mathrm{B}T)$, with $k_\mathrm{B}$ the Boltzmann constant and $T$ the temperature.
The Bogoliubov dispersion relation in the local-density approximation is 
\begin{equation}\label{eq:dispersion}
    \epsilon_\textbf{k}(\textbf{r})=\sqrt{\tau_\textbf{k}(\tau_\textbf{k}+2\tilde{U}(\textbf{k})|\Psi(\textbf{r})|^2)},
\end{equation}
with $\tilde{U}(\textbf{k})$ the Fourier transform of the interaction potential~\eqref{eq:Uint},
\begin{equation}
    \tilde{U}(\textbf{k})= g+\frac{\mu_0\mu_m^2}{3}\left(3\cos^2\theta_\textbf{k}-1\right),
\end{equation}
where $\theta_\textbf{k}$ is the angle between $\textbf{k}$ and the dipole polarization axis.

For $\edd>1$, Eq.~\eqref{eq:dispersion} becomes imaginary at low momenta and yields imaginary contributions to the fluctuations terms of Eqs.~\eqref{eq:HLHY} and~\eqref{eq:HTherm}, with a stronger relative weight in the latter case~\cite{Waechtler2016Quantum}. 
While the imaginary contribution can be neglected in Eq.~\eqref{eq:HLHY},  extra care should be taken in evaluating the thermal fluctuation term in this regime.
Rather than selecting a momentum cutoff, as in some previous studies of dipolar supersolids in a thermal setting~\cite{SanchezBaena2023Heating,SanchezBaena2024SuperfluidSupersolid,He2025Accessing}, we instead consider an ansatz for $H_{\mathrm{T}}$ (see also Refs.~\cite{Aybar2019Temperature,Yu2025Thermally} for a similar approach with an alternative functional form) that well-approximates the density dependence, namely
\begin{equation}
    H_{\mathrm{T}}\approx A(a_s, T)\mathrm{e}^{-B(a_s, T)|\Psi(\textbf{r})|}\, ,
    \label{eq:HThermApprox}
\end{equation}
where $A(a_s, T)$ and $B(a_s, T)$ are best fit to match the exact value of the integral within the mean-field stable regime; see App.~\ref{app:surfaces} for the exact form of $A$ and $B$.

\begin{figure}
    \centering
    \includegraphics[width=1.0\linewidth]{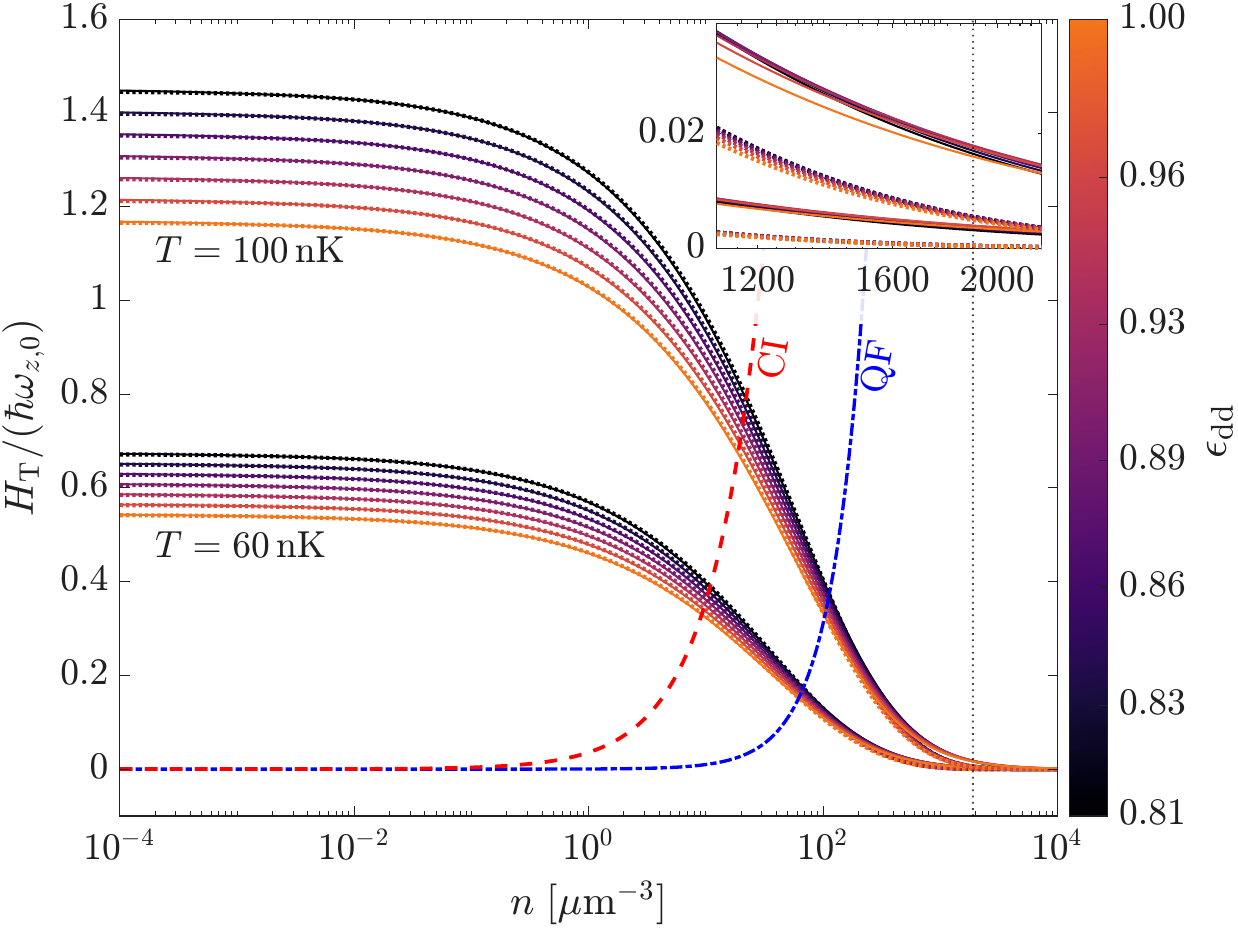}
    \caption{\textbf{Thermal fluctuations.} Effective thermal potential as a function of the local 3D density $n=|\Psi|^2$. For the two temperatures $T = 60$ and $\SI{100}{\nano\K}$, solid lines show Eq.~\eqref{eq:HTherm} for a range of $\edd$ (see colorbar) within the stable region. We also show our approximate model [Eq.~\eqref{eq:HThermApprox}] with best-fit parameters $A$ and $B$ in the same colors as the dotted lines. The exact integral and model lie nearly on top of one another, with a slight underestimation at very high densities, highlighted in the inset. The contact interaction $g|\Psi|^2$ (CI, red dashed) and quantum fluctuation $\gamma_{\mathrm{Q}}|\Psi|^3$ (QF, blue dot-dashed) are plotted for comparison at $\edd=1$. The maximum 3D density encountered across the states we consider in this paper is indicated as a vertical dotted line. The energies are given in units of $\hbar\omega_{z,0}$, with reference frequency $\omega_{z,0} = 2\pi \times\SI{150}{\Hz}$.}
    \label{fig:HTerm}
\end{figure}

In Fig.~\ref{fig:HTerm}, we show numerical solutions of Eq.~\eqref{eq:HTherm} at two temperatures for a range of relative interaction strengths $\edd$ in a log-linear plot for easier visualization across a wide density range.
The general trend is that thermal effects yield an energy increase that tends to dominate at low densities, while becoming negligible at high densities.
This behavior encourages the formation of high-density regions, i.e., droplets or stripes, with the effect becoming stronger at higher temperatures.
Our model of Eq.~\eqref{eq:HThermApprox} is also shown in Fig.~\ref{fig:HTerm} (dashed lines) and well-captures the density dependence of Eq.~\eqref{eq:HTherm} over many orders of magnitude, with a slight deviation at very high densities; see inset.
We note that in this region where the theories deviate slightly, the relative difference is negligible compared to the dominant energy scales given by the contact ($\approx$ dipolar when $\edd=1$) and quantum fluctuation contributions, also shown in Fig.~\ref{fig:HTerm}.
The maximum density of the states we encounter in our simulations is approximately $1.9\times10^{3}~\um^{-3}$, denoted as a vertical dotted line in Fig.~\ref{fig:HTerm}. 

In the following, we solve Eq.~\eqref{eq:TeGPE} in imaginary time to find the ground (and metastable) states of the system; see App.~\ref{app:numerics} for more details.

\section{Phase diagram}\label{sec:phasediagram}
\begin{figure}[t!]
    \centering
    \includegraphics[width=1\linewidth]{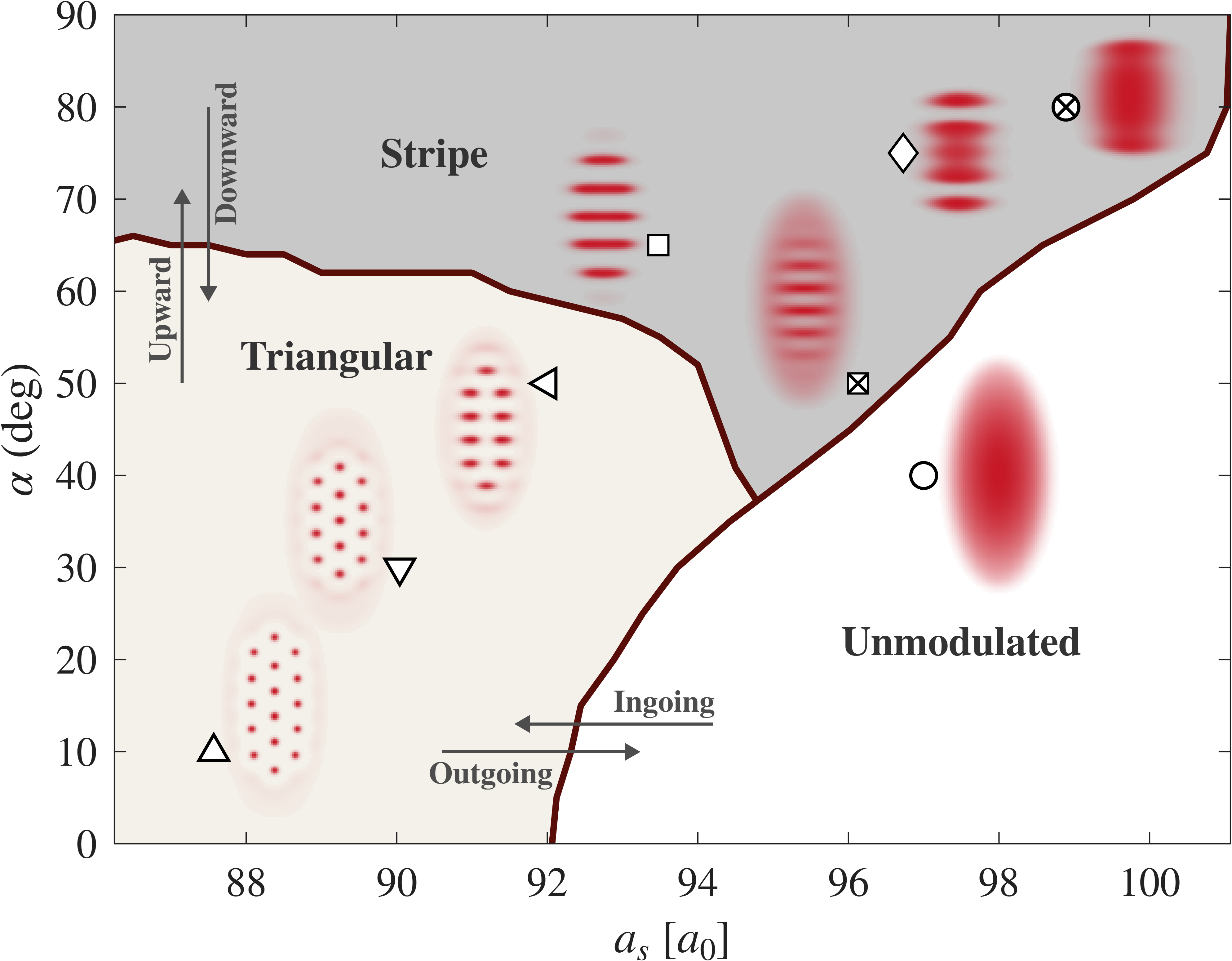}
    \caption{\textbf{Exemplary phase diagram.} Phase diagram of a \Dyb\ gas in a surfboard-shaped trap for the reference parameters $N=2\times10^5$, $T=\SI{60}{\nano\K}$ and $\nu_z = \SI{150}{\Hz}$ and varying dipole tilt $\alpha$ and scattering length $\as$. The insets show exemplary density distributions (cut at $z=0$) of the ground states at the parameters of the corresponding symbols. Unmodulated, triangular, and stripe phases (shaded regions) are identified based on the structure of the density profile; see App.~\ref{app:classification}. Arrows indicate directions of transition crossings used in Sec.~\ref{subsec:unmod_transition} and~\ref{subsec:stripe_transition}: ingoing/outgoing for the unmodulated-to-modulated transition and upward/downward for the triangular-to-stripe transition.}
    \label{fig:phasediagram}
\end{figure}

We calculate the phase diagram as a function of dipole tilt $\alpha$ and contact interaction strength $\as$, and investigate the influence of particle number $N$, vertical trapping frequency $\nu_z$, and temperature $T$ on the underlying phase transitions.
One set of parameters is chosen as the reference, with $N=2\times 10^5$, $T=\SI{60}{\nano\K}$ and $\nu_z = \SI{150}{\Hz}$, which are experimentally realistic values~\cite{Chandrashekara2026Competing}.
The phase diagram for these parameters is shown in Fig.~\ref{fig:phasediagram}; see Sec.~\ref{subsec:unmod_transition} and Sec.~\ref{subsec:stripe_transition} for details on the state classification and the phase boundaries. 
The insets show exemplary states, represented by $n(x,y, z=0)$ with $n(\textbf{r})=\left| \Psi(\textbf{r})\right|^2$ being the real-space density distribution. 

As already found in previous works~\cite{Ripley2023Twodimensional, Hertkorn2021Pattern,Zhang2024Metastable,Lima2025Supersolid,SanchezBaena2025Tilted, Chandrashekara2026Competing}, at large $a_s$, the system is in the unmodulated superfluid phase ($\bigcirc$) for all tilt angles $\alpha$, whereas, at low $a_s$, it forms density-modulated states of variable crystal structures.
The unmodulated-to-modulated transition is studied in Sec.~\ref{subsec:unmod_transition} using the contrast of the density modulation as an order parameter for all crystal structures~\cite{Zhang2019Supersolidity,Biagioni2022Dimensional}. 
The crystal structure primarily varies with the dipole tilt $\alpha$.   
For low tilt angles and scattering lengths, the system forms a triangular lattice of droplets ($\triangle$ and $\triangledown$). 
For intermediate $\alpha$, many lattice configurations appear to have nearly equal energies and form metastable states. 
In particular, states where droplets are aligned along the $x$ axis can become the ground state ($\triangleleft$).
Starting from this configuration, stripe states can be reached by increasing either the scattering length ($\boxtimes$) or the tilt angle ($\square$). 
The latter case is studied in detail in Sec.~\ref{subsec:stripe_transition}. 
For the former case ($\triangleleft \rightarrow  \boxtimes $), we observe that increasing $\as$ yields a density modulation with decreasing contrast along $x$, which ultimately results in stripes when the modulation vanishes. 
For $\alpha \gtrsim 60^\circ$ at high scattering lengths, the stripe states show different structures for increasing scattering lengths: the states undergo two transitions, where first the inner stripes merge ($\Diamond$) and then, in a second step, the outer stripes follow ($\otimes$). This regime is discussed in Sec.~\ref{subsec:large_tilts}.

\subsection{Unmodulated-to-modulated transition}\label{subsec:unmod_transition}

\subsubsection{Transition characterization} \label{subsubsec:transition}
\begin{figure}[t!]
    \centering
    \includegraphics[width=1\linewidth]{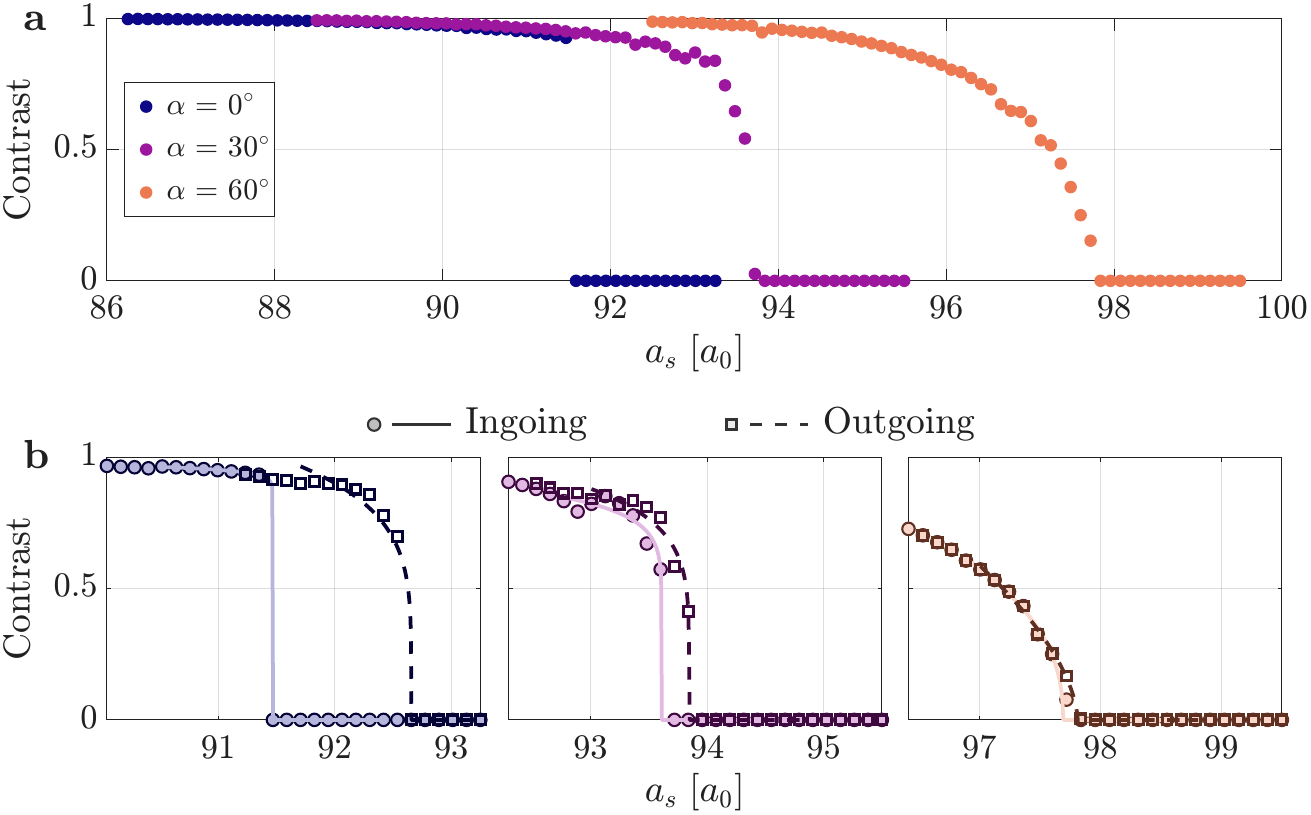}
    \caption{\textbf{Density contrast curves and hysteresis analysis.} Density contrast of the simulated states as a function of $a_s$ for the reference parameters $N = 2\times 10^5$, $T = \SI{60}{\nano \K}$ and $\nu_z = \SI{150}{\Hz}$. (a) Density contrast for three different tilt angles $\alpha = 0^\circ$ (blue), $30^\circ$ (violet), and $60^\circ$ (orange) with a spacing of $\delta a_s \approx 0.12a_0$. Panel~(b) shows the contrast curves for the ingoing (circle) and outgoing (square) directions for $\alpha = 0^\circ$, $30^\circ$ and $60^\circ$ (from left to right), as well as the corresponding fits (solid and dashed lines, respectively) to Eq.~\eqref{eq:Landau} in the region of $1a_0$ below the transition for both directions.}
    \label{fig:contrasts}
\end{figure}

Building on previous Landau analysis in the thermodynamic limit~\cite{Zhang2019Supersolidity,Biagioni2022Dimensional}, we  use the density modulation contrast as an order parameter for the unmodulated-to-modulated transition. 
In our trapped system, we calculate the contrast from density profiles along triangulation edges between density structures. 
We calculate a contrast along each of these edges and take the average over all edge contrasts; see App.~\ref{app:contrast} for more details. 
Fig.~\ref{fig:contrasts}~(a) shows the resulting contrast as a function of $\as$ for different dipole tilts and for our reference settings. The contrast remains zero in the unmodulated phase and becomes finite in the modulated regime. It exhibits large, discontinuous jumps for low $\alpha$ while it evolves smoothly with $\as$ for larger $\alpha$. 
This suggests a first-order (second-order) transition at small (large) dipole tilts.

To further characterize the order of the unmodulated-to-modulated transition, we aim to quantify its degree of discontinuity. However, this cannot be reliably measured through the magnitude of the contrast jump.
Indeed, its exact value depends on the thresholds chosen for identifying peaks, which are set once for the full parameter range. 
This is also responsible for the small jumps and fluctuations observed in Fig.~\ref{fig:contrasts}~(a). 
Moreover, the energy landscape of this system may host numerous local minima, close in energy, especially in the vicinity of the phase transition, and the state reached by imaginary-time propagation can depend on its initialization. In particular, different initial noise realizations were found to lead under imaginary-time propagation to states with different arrangements or symmetries at otherwise identical parameters. This makes it difficult to unambiguously identify the precise structural arrangement corresponding to the true global minimum and thereby calculate its density contrast. To circumvent those ambiguities, we characterize the degree of continuity of the underlying phase transitions through imaginary-time hysteresis analysis~\cite{Boettcher2019Transient}. 

Imaginary-time hysteresis analysis relies on a bidirectional parameter sweep across the phase boundary while the system is evolved according to the TeGPE [Eq.~\eqref{eq:TeGPE}] in imaginary time \cite{Boettcher2019Transient}. More specifically, we initialize the state through imaginary-time propagation using a scattering-length value deep in the unmodulated (modulated) regime and move across the transition by decreasing (increasing) $\as$ in small steps while using the last simulation result as the initial state for the next simulation at a slightly lower (higher) $\as$. 
In the following, we refer to this scheme as the transition crossing in the ingoing (outgoing) direction; see also Fig.~\ref{fig:phasediagram}. 

Through the parameter sweeps in imaginary time, these schemes probe the metastability of the phase in which they are initialized: as long as the initial phase remains metastable, the imaginary time evolution results in a state of similar structure and therefore of similar contrast. If the ingoing and outgoing schemes yield the same contrast curves, no hysteresis is observed, no metastability is implied, and the underlying transition is continuous. Conversely, if the curves differ, different transition points are predicted depending on the sweep direction, and the interval between these points corresponds to a region of bistability, in which both phases persist as metastable states; the underlying transition is discontinuous.

Fig.~\ref{fig:contrasts}~(b) shows the density contrast curves in the transition crossing in the ingoing (circles) and outgoing (squares) directions for the same parameters as Fig.~\ref{fig:contrasts}~(a). 
At low tilts, hysteresis between in- and outgoing directions is observed, with the corresponding critical scattering lengths, $a_c$, at which modulation emerges, being shifted between the two directions, taking lower values for the ingoing direction. 
This confirms the discontinuous character of the transition, suggested by the jump in Fig.~\ref{fig:contrasts}~(a), at low tilts. 
Instead, at larger tilts, the contrast curves of in- and outgoing directions coincide, yielding a vanishing hysteresis, indicative of a continuous transition as suggested by the absence of a jump in Fig.~\ref{fig:contrasts}~(a).

To quantitatively describe the evolution of the density contrast across the transition, we use a phenomenological power-law form inspired by Landau theory,  
\begin{align}\label{eq:Landau}
     C(a_s) = \begin{cases}
        A\left( a_c - a_s\right)^\beta & a_s \leq a_c \\
        0 & \, \text{else,}
        \end{cases}
\end{align}
with fit parameters $A$, $a_c$, and $\beta$. We fit Eq.~\eqref{eq:Landau} to the contrast curves in a range extending from the highest $\as$ simulated down to $1\,a_0$ below the first non-zero contrast value encountered upon decreasing $\as$.
Exemplary fits are shown in Fig.~\ref{fig:contrasts}(b) for the ingoing (outgoing) direction [solid (dashed) lines]. 
The functional form of Eq.~\eqref{eq:Landau} is chosen to encompass the variety of behaviors envisioned at the transition: For a second-order transition, mean-field Landau theory gives $\beta = 1/2$ in the thermodynamic limit, while this exponent can vary depending on the transition universality class, and the scaling behavior can be modified by finite size effects. 
Conversely, $\beta\rightarrow 0$ produces an abrupt onset of the contrast and can therefore phenomenologically capture the discontinuous behavior expected at a first-order transition, although the critical scaling form of Eq.~\eqref{eq:Landau} does not strictly apply in this case.  
In the following, we will not comment on the value taken by the exponent $\beta$ and simply use the fit to Eq.~\eqref{eq:Landau} to extract $a_c$. 
We characterize the bistability range around the transition by 
\begin{align}\label{eq:hysteresis_shift}
    \Delta a_c = a_c^\mathrm{out} - a_c^\mathrm{in},
\end{align}
where $a_c^\mathrm{out}$ and $a_c^\mathrm{in}$ are the critical scattering lengths extracted from fits to the outgoing- and ingoing-direction curves, respectively.
Note that the phase diagram in Fig.~\ref{fig:phasediagram} shows the average phase boundary for the two directions.

\begin{figure}
    \centering
    \includegraphics[width=1\linewidth]{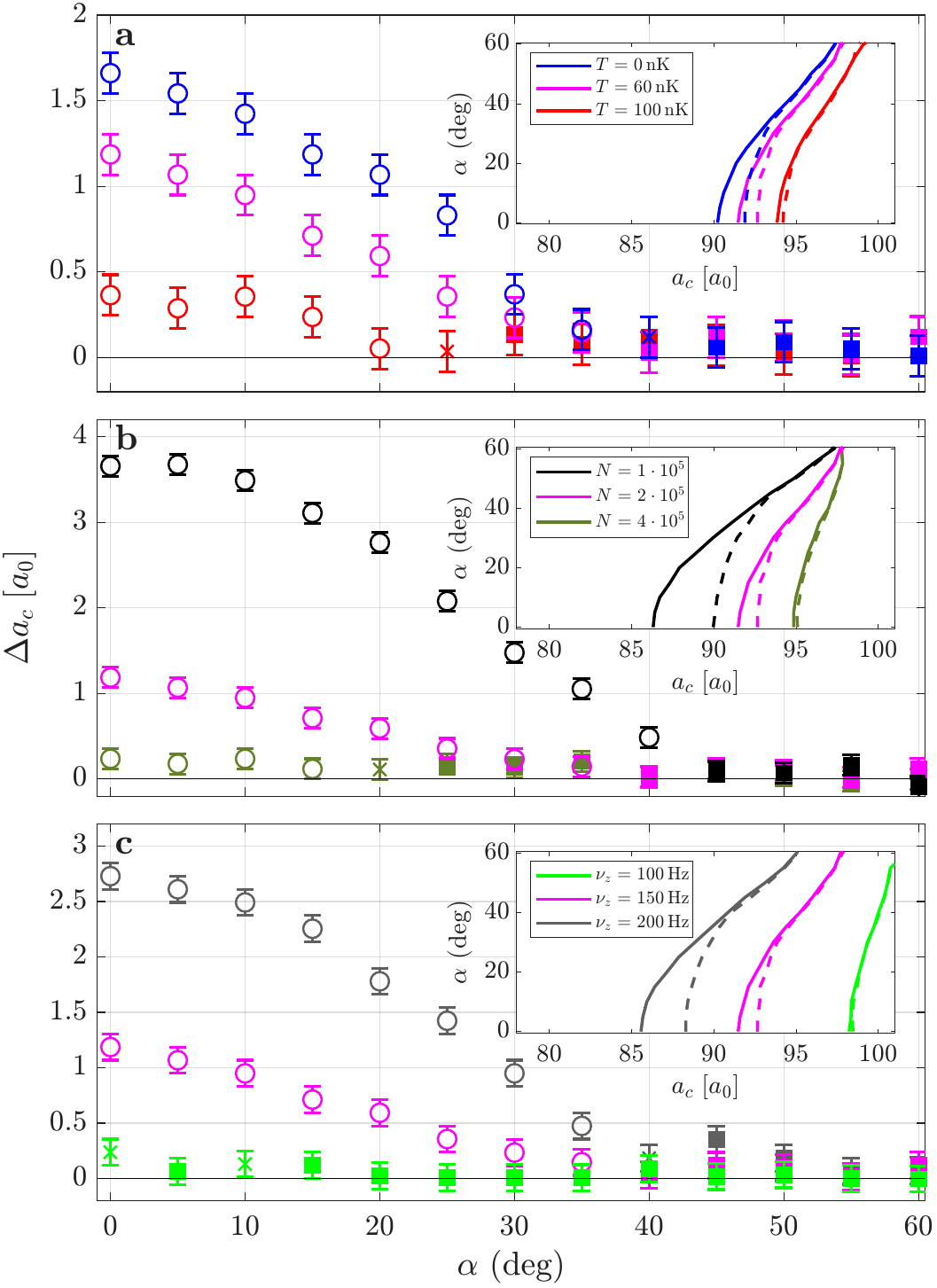}
    \caption{\textbf{Hysteresis and phase boundaries of the unmodulated-to-modulated transition}. Each panel shows the bistability range $\Delta a_c = a_c^\mathrm{out} - a_c^\mathrm{in}$ (symbol) and, in the inset,  the critical scattering lengths for the ingoing, $a_c^\mathrm{in}$ (solid), and outgoing, $a_c^\mathrm{out}$ (dashed), directions. The error bars for $\Delta a_c$ correspond to the resolution in $a_s$, given by $\delta a_s \approx 0.12a_0$. Panel~(a) shows various temperatures $T \in \{0, 60, 100\}\,\mathrm{nK}$, (b) particle numbers $N \in \{1, 2, 4\}\times 10^5$, and (c) vertical trapping frequencies $\nu_z \in \{100, 150, 200\}\, \mathrm{Hz}$. If not varied, $T=\SI{60}{\nano\K}$, $N = 2\times 10^5$, and $\nu_z = \SI{150}{\Hz}$ are the reference parameters, always represented by the magenta curves. Open circles correspond to triangular states, crosses to states where either the ingoing or outgoing direction found stripe states, and  filled squares represent stripe states found for both directions.}
    \label{fig:hysteresis}
\end{figure}

\subsubsection{Transition character for varying system parameters}\label{subsubsec:variations}

Fig.~\ref{fig:hysteresis} shows the bistability range $\Delta a_c$ as a function of the dipole tilt $\alpha$ and the phase boundaries (insets) for different temperatures~(a), particle numbers~(b), and vertical trapping frequencies~(c). The uncertainty in the critical scattering length (error bars in Fig.~\ref{fig:hysteresis}) is set by the chosen simulation steps $\delta a_s\approx0.12a_0$ along the transition crossings. This is smaller than the typical experimental resolution in current magnetic-atom experiments; see e.g.~Refs.~\cite{Chomaz2022Dipolar,Chomaz2026Quantum,Chomaz2019LongLived,Tanzi2019Observation,Boettcher2019Transient,Norcia2021Twodimensional,Chandrashekara2026Competing}.  

For all choices of gas parameters, we observe a similar general trend: The bistability range is the largest at zero tilt and smoothly decreases with increasing tilt until it vanishes at $\alpha=\alpha_c$ and a continuous transition is found at larger tilts $\alpha_c\leq \alpha\lesssim 60^\circ$. 
A similar influence of the tilting angle $\alpha$ was also reported in Refs.~\cite{Lima2025Supersolid, SanchezBaena2025Tilted}. 
The transition becomes effectively discontinuous again at large angles $\alpha\gtrsim 60^\circ$ as hysteresis reoccurs; see Sec.~\ref{subsec:large_tilts}. 
The gas parameters, however, modify the strength of the discontinuity as measured by the bistability range, as well as the angle $\alpha_c$ at which the transition becomes continuous, as detailed below. 

First, increasing temperature leads to a smaller bistability range $\Delta a_c$ for all $\alpha$ and a continuous transition occurring at a smaller angle $\alpha_c$; see Fig.~\ref{fig:hysteresis}(a). 
It also shifts the phase boundary to larger scattering lengths; see Fig.~\ref{fig:hysteresis}(a, inset). 
This is similar to previous observations~\cite{SanchezBaena2023Heating,He2025Accessing,Bombin2025Creating} and can be explained by the fact that the thermal Hamiltonian~[Eq.~\eqref{eq:HTherm}], which tends to bias the system towards high-density droplets, increases with temperature; see Fig.~\ref{fig:HTerm}.

Second, increasing the particle number~$N$ also results in a decrease of $\Delta a_c$ for all $\alpha$, a shift of $\alpha_c$ to lower values, and a shift of the phase boundary to larger scattering lengths; see Fig.~\ref{fig:hysteresis}(b). These behaviors are consistent with previous studies. 
In particular, calculations in the thermodynamic limit have shown that, starting from low densities, increasing the density raises the critical scattering length and narrows the bistability region, with the transition eventually changing from first to second order upon entering the stripe phase~\cite{Blakie2020Supersolidity,Zhang2019Supersolidity, Biagioni2022Dimensional,Lima2025Supersolid, Ripley2023Twodimensional}. 
Similar trends have been predicted for increasing particle number in untilted trapped systems~\cite{Biagioni2022Dimensional, Shi2026MatterWave}. 

Finally, the vertical confinement is found to have a strong impact on the continuity and position of the transition [Fig.~\ref{fig:hysteresis}~(c)], while, to the best of our knowledge, the effect of this parameter has not been systematically studied so far. 
By lowering $\nu_z$, the transition moves to larger scattering lengths (see inset) and becomes more continuous with $\Delta a_c$ decreasing for all $\alpha$ and $\alpha_c$ shifting to lower values. Remarkably, for the lowest $\nu_z$ considered, the transition appears to be continuous even at $\alpha=0$. 

Based on previous studies showing that the dimensionality of the formed crystal dictates the order of the unmodulated-to-modulated transition~\cite{Zhang2019Supersolidity, Biagioni2022Dimensional,Lu2015}, see also Sec.~\ref{sec:intro}, we investigate the crystal structure in the region of $1a_0$ below the transition.
To do so, we consider the state's power spectrum, which is the absolute norm square of the Fourier transform of the real-space density, and extract the parameter $R$, the ratio of the two largest peaks in the upper sector of the power spectrum; see Sec.~\ref{subsec:stripe_transition}, App.~\ref{app:classification}, and Ref.~\cite{Chandrashekara2026Competing} for more details. 
Triangular states have two (or three, depending on the orientation) similarly strong peaks in the upper sectors, yielding $R\approx 1$, while stripes have one dominant peak, yielding $R\approx 0$.  
States with $R \geq 0.05$ are identified as triangular states and are marked in Fig.~\ref{fig:hysteresis} by open circles. 
Cases where ingoing and outgoing directions give different structural classifications are marked by crosses, and states where both directions give stripes are marked by filled squares.

Vanishing hysteresis effects are often related to the appearance of stripe states, and vice versa. Yet, this correspondence is not systematic. Continuous transition may occur towards two-dimensional crystal structures; see e.g., Fig.~\ref{fig:hysteresis}(a), $T=\SI{100}{\nano\K}$, $\alpha=20^\circ$, or Fig.~\ref{fig:hysteresis}(b), $N=4\cdot 10^5$, $\alpha=15^\circ$ ; while stripe states may be formed through (weakly) discontinuous phase transitions, see e.g. Fig.~\ref{fig:hysteresis}(a), $T=100\,$nK, $\alpha=30^\circ$, or Fig.~\ref{fig:hysteresis}(b), $N=4\cdot 10^5$, $\alpha=25^\circ$. We note that the presence of the trap and its inhomogeneity seem to play an important role. In such a trap, the density modulation appears first dominantly at the trap center with side modulation of variable strength and geometries; see App.~\ref{app:classification} for examples. Depending on the parameters, the central crystal structures may form a one-dimensional pattern, while the side modulation can appear either as part of the crystal, making it two-dimensional, or instead as forming a diffuse halo, not directly contributing to the crystal dimensionality.  This distinction is not unambiguously defined, neither by eye nor through the power spectrum analysis. 
We note that, by comparing the present results to those obtained in a pancake-shaped trap, we found no significant influence of the anisotropy of the surfboard trap on the structural arrangements and hysteresis; see App.~\ref{app:pancake_trap}.

\subsection{Triangular-to-stripe transition}\label{subsec:stripe_transition}
To investigate the triangular-to-stripe phase boundary, we consider the previously introduced $R$ parameter, which was suggested as an order parameter for the triangular-to-stripe transition in Ref.~\cite{Chandrashekara2026Competing}. Based on $R$, we classify the crystalline structures as stripe or triangular arrangements and investigate hysteresis effects upon tuning $\alpha$ at fixed $\as$. That is, as in Sec.~\ref{subsec:unmod_transition}, we perform two complementary sets of imaginary time evolutions: an upward (downward) set, which we initialize at low (large) $\alpha$ in the triangular (stripe) phase and increase (decrease) $\alpha$ by recursively taking the last ground state as the next initial state; see also Fig.~\ref{fig:phasediagram}. 

\begin{figure}
    \centering
    \includegraphics[width=1\linewidth]{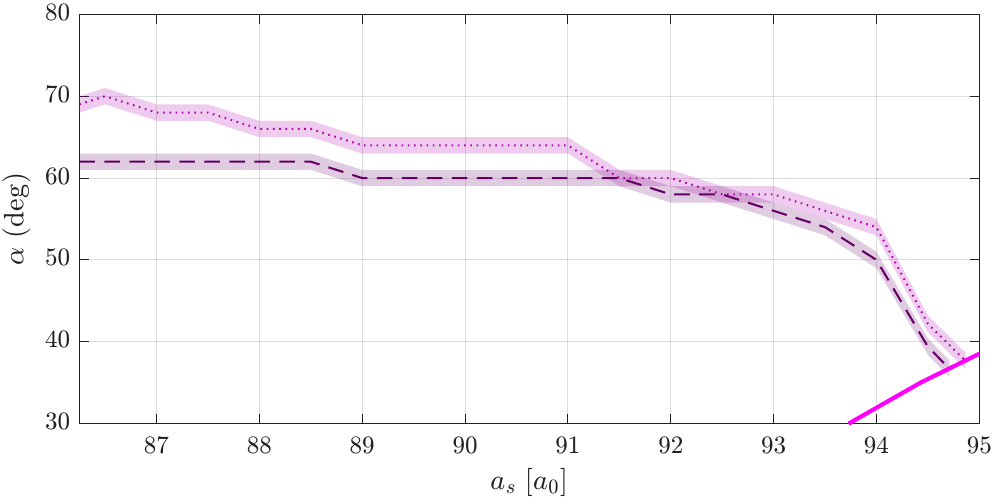}
    \caption{\textbf{Hysteresis at the triangular-to-stripe transition.} Phase boundaries for the transition crossing in the upward (dotted) and downward (dashed) directions for the reference parameters $N = 2\times 10^5$, $T = \SI{60}{\nano \K}$ and $\nu_z = \SI{150}{\Hz}$. The boundaries are obtained by initializing the imaginary time in the triangular and stripe phases and then recursively moving up and down in $\alpha$ at a fixed $a_s$, respectively. The uncertainty of the boundaries is expressed by the bands with a width of $\delta\alpha = 2^\circ$. The solid line shows the unmodulated-to-modulated phase boundary; see Sec.~\ref{subsec:unmod_transition}.}
    \label{fig:both_phase_boundaries}
\end{figure}

Fig.~\ref{fig:both_phase_boundaries} shows the corresponding phase boundaries for our reference parameters and for the upward and downward crossings of the triangular-to-stripe phase transition (dashed-dotted and dashed lines, respectively), together with the unmodulated-to-modulated phase boundary as in Fig.~\ref{fig:phasediagram}. At low scattering lengths, 
a bistability region of a few degrees is found. This region shrinks and nearly disappears while increasing $\as$, indicating a more continuous transition for $\as$ approaching the modulated-to-unmodulated transition. Due to the previously highlighted ambiguity of the crystal structure identification in our finite-size system, especially close to the modulated-to-unmodulated transition (see Sec.~\ref{subsec:unmod_transition}), the structural bistability can only be assessed qualitatively. By repeating similar simulations for different gas parameters, no clear dependence of the structural bistability in $\alpha$ on $T$, $N$, and $\nu_z$ is found.

Note that the triangular-to-stripe phase boundary in Fig.~\ref{fig:phasediagram} corresponds to the average phase boundary for the two directions.  

\subsection{States at large tilt angles}\label{subsec:large_tilts}

The states at large tilt angles exhibit distinct shapes (see Fig.~\ref{fig:phasediagram}), which we here investigate further, in relation to the underlying unmodulated-to-modulated transition.

\begin{figure}
    \centering
    \includegraphics[width=1\linewidth]{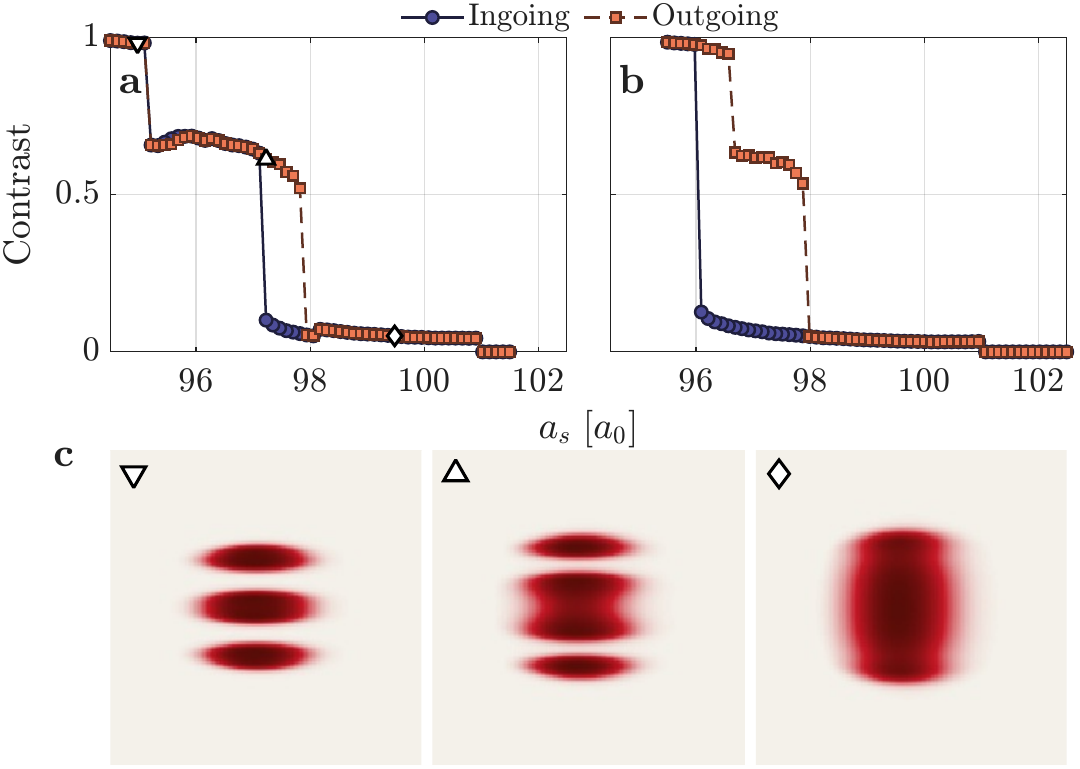}
    \caption{\textbf{Hysteresis at large tilt angles.}  Contrast curves with connections between data points for a guide for the eye for ingoing (circles, solid) and outgoing (squares, dashed) are shown for the reference parameters $N = 2\times 10^5$, $T = \SI{60}{\nano \K}$ and $\nu_z = \SI{150}{\Hz}$, and for $\alpha = 80^\circ$ and $90^\circ$ in (a) and (b), respectively. (c) Density profiles of the simulated states in the outgoing direction for $\alpha = 80^\circ$ and at the parameters indicated by the corresponding symbols, $\triangledown, \triangle$ and $\Diamond$, in~(a).}
    \label{fig:large_tilts}
\end{figure}

Fig.~\ref{fig:large_tilts} shows the density contrast curves as a function of $\as$ for the reference parameters for both the ingoing (circles, solid) and outgoing (squares, dashed) directions at $\alpha = 80^\circ$ (a) and $90^\circ$ (b).
Jumps in the density contrast and hysteresis effects are found, indicative of discontinuous transitions.
We note that hints of such a discontinuity have also been found for infinite systems~\cite{Blakieprivate}. 
Furthermore, the transition seems to happen in distinct stages, marked by successive jumps of the density contrast; see also~\cite{Sanchez2026Quasi}. These seem to connect to rearrangements specific to finite-size systems, as illustrated in the density profiles of Fig.~\ref{fig:large_tilts}(c) for $\alpha = 80^\circ$. The different profiles are separated by contrast jumps [see highlighted symbols Fig.~\ref{fig:large_tilts}(a)], and we note that neighboring states with similar contrast exhibit similar density profiles. These different states are all modulated along $y$, yet they exhibit different crystal structures. At low $\as$, three stripe-like structures are observed ($\triangledown$). 
Upon increasing $\as$, the central 'stripe' seems to extend and get modulated along $y$ ($\triangle$), until it merges with the side 'stripes' while retaining a slight density modulation ($\Diamond$).

We note that the exact behavior changes depending on the chosen parameters (compare $\alpha = 80^\circ$ and $90^\circ$) and on the direction of the transition crossing (compare ingoing and outgoing directions at $\alpha = 90^\circ$).

\section{Conclusion}\label{sec:conclusion}

In conclusion, we have investigated the phase diagram of a dipolar quantum gas confined in a surfboard-shaped trap over a range of experimentally relevant parameters, tuning both the dipole orientation and the contact interaction strength. The resulting phase diagram encompasses a rich variety of phases, from unmodulated superfluids to supersolid triangular droplet arrays and supersolid stripe states, separated by phase transitions of different types.   

We have explored the unmodulated-to-modulated transition, which shows the largest bistability region at $\alpha =0$, and becomes smoothly more continuous upon increasing $\alpha$ up to $\alpha \approx 60^\circ$. Typically, but not systematically, the emergence of a continuous transition relates to the density-modulated state becoming of stripe character. Increasing particle number or temperature or decreasing the vertical trapping frequency shrinks the bistability region and results in a continuous transition at lower $\alpha$. 
For $\alpha \gtrsim 60^\circ$, the transition becomes discontinuous again, and finite size effects seem to play a significant role. We have also explored the triangular-to-stripe transition, which appears discontinuous at small $\as$ and increasingly continuous upon approaching the unmodulated-to-modulated transition.  

The present results open several directions for further investigating the nature of phase transitions of the dipolar supersolid phase diagram. First, a detailed comparison of our results with predictions for infinite systems could help disentangle intrinsic properties of the underlying phase transitions from finite-size and inhomogeneity effects introduced by the trap and establish how the different regimes identified here connect to their thermodynamic-limit counterparts. Furthermore, it would be interesting to explore further different geometries and, in particular, to go beyond simple harmonic trapping and explore how different spatial inhomogeneities affect the order and location of the transitions.

Beyond the equilibrium properties considered here, an important question is how the predicted transition order manifests itself under experimentally realistic dynamical protocols. Real-time simulations, including thermal and stochastic fluctuations, could help understand to what extent the equilibrium metastability and hysteresis persist on experimental time scales and identify dynamical signatures of the transition order, e.g., by studying quenches of the control parameters. Furthermore, such dynamical studies could provide access to universal phenomena associated with the crossing of phase transitions, including Kibble–Zurek mechanism across continuous transitions~\cite{Kibble1976Topology,Zurek1985Cosmological} and nucleation-driven false-vacuum decay across first-order transitions~\cite{Callan1977Fate,Coleman1977Fate}. First theoretical studies of these mechanisms in dipolar supersolids have been reported~\cite{Kirkby2025Kibble,Kirkby2026Morphological, Wong2026SpinodalControlled}. Systematically exploring these dynamics across different regions of the phase diagram could reveal how the dynamical properties of the transitions, including their scaling laws and universal behavior, evolve as their order changes with the system parameters. 

Finally, the structural transition between triangular and stripe states calls for a more complete theoretical description. Developing an order parameter that remains well defined in finite trapped systems, together with an effective Landau theory incorporating the competing crystalline orders, could provide a quantitative framework for characterizing this transition and elucidating the mechanisms that determine its order.

\begin{acknowledgments}
        The authors thank K. Chandrashekara, J. Gao, C. G\"olzh\"auser, B. Lanigan, L. Platt, L. Schulz, B. Blakie, and D. Baillie for helpful discussions and collaboration on related work. They acknowledge support by the European Research Council (ERC) under the European Union’s Horizon Europe research and innovation program under grant number 101040688 (project 2DDip), by the Deutsche Forschungsgemeinschaft (DFG, German Research Foundation), through SFB 1225 ISOQUANT (Project-ID 273811115), the ANR-DFG  bilateral project DISQUTT (Project-ID 530000649) and under Germany's Excellence Strategy -- EXC 2181/1 -- 390900948 (the Heidelberg STRUCTURES Excellence Cluster), and by the state of Baden-W{\"u}rttemberg through bwHPC and the German Research Foundation (DFG) through grant INST 35/1597-1 FUGG. W.K. acknowledges the support of the Natural Sciences and Engineering Research Council of Canada (NSERC), [funding reference number PDF-577924-2023]. Views and opinions expressed are however those of the authors only and do not necessarily reflect those of the European Union or the European Research Council. Neither the European Union nor the granting authority can be held responsible for them.  
    \end{acknowledgments}
 
\appendix

\section{Finite-temperature surfaces}\label{app:surfaces}
\begin{figure}
    \centering
    \includegraphics[width=1.0\linewidth]{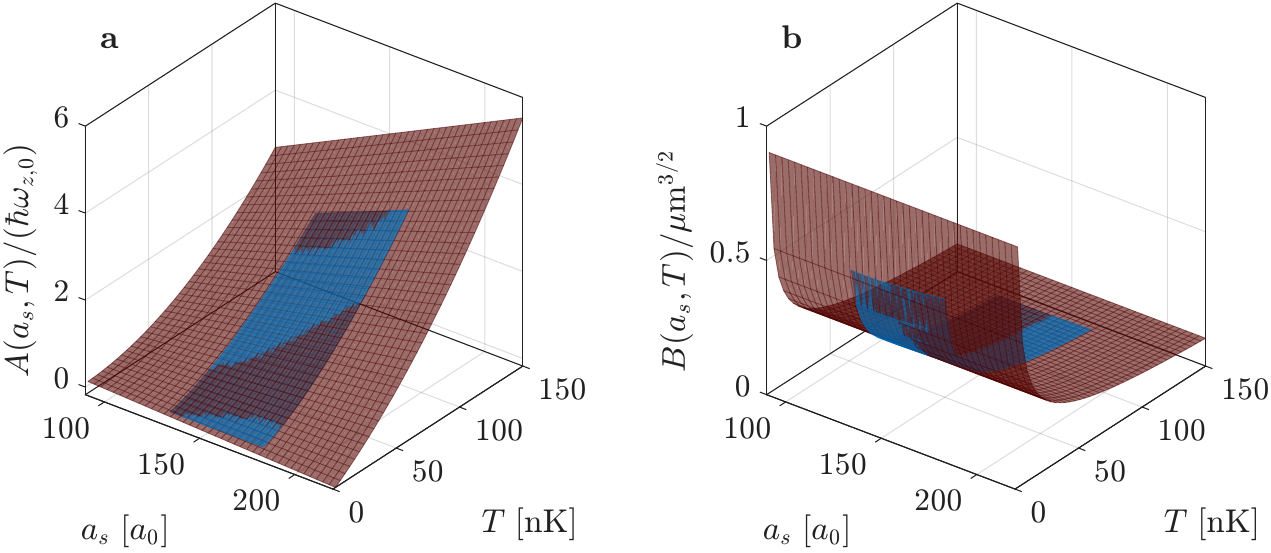}
    \caption{\textbf{Finite-temperature surfaces.} Coefficients of Eq.~\eqref{eq:HThermApprox} in the ($a_s$,$T$) landscape described by Eq.~\eqref{eq:surfaceA} in (a) and Eq.~\eqref{eq:surfaceB} in (b). The coefficients in the stable region are shown in blue, and the extrapolations into the mean-field unstable region are shown in dark red.}
    \label{fig:surfaces}
\end{figure}

Here, we demonstrate how the thermal contribution given in Eq.~\eqref{eq:HThermApprox} is extended to the mean-field-unstable region. 
First, recall that the integral given in Eq.~\eqref{eq:HTherm} converges when $\edd<1$, we proceed by evaluating this integral numerically within the stable regime.
The exponential form given in Eq.~\eqref{eq:HThermApprox}, with best-fit parameters $A$ and $B$, then provides a good approximation of the functional dependence of the exact density dependence, as can be seen in Fig.~\ref{fig:HTerm}.
To extend beyond the stable region, we assume that the general effect of temperature as a function of density does not abruptly change at $\edd=1$ but rather that $H_{\mathrm{T}}[a_s,T,n(\mathrm{r})]$ is a smooth functional across our parameter space.
To this end, we take $A(a_s,T)$ and $B(a_s,T)$ to be smooth functions of temperature and scattering length and fit them to the following functions in the stable region ($\edd<1$),
\begin{align}
    A(a_s, T) &= X_{00} + X_{10}a_s + X_{01}T + X_{20}a_s^2 \nonumber \\ & \quad + X_{11}a_sT + X_{02}T^2 \label{eq:surfaceA} \\
    B(a_s,T) & = \frac{X_1}{X_2T + X_3\sqrt{T} + X_4} + X_5\mathrm{e}^{-X_6a_s} .\label{eq:surfaceB}
\end{align}
The choice for these functions was determined by examining the behavior of $A$ and $B$ in the stable region and selecting functions that well-capture the observed behavior without overfitting.
The fits of Eqs.~\eqref{eq:surfaceA}~and~\eqref{eq:surfaceB} are shown in Fig.~\ref{fig:surfaces}, as panels (a) and (b), respectively. 
The blue areas show the coefficient values as calculated via the integral in Eq.~\eqref{eq:HTherm} in the window $T \in [5, 120] \, \mathrm{nK}$ and $a_s \in [131, 180]\, a_0$ in the stable region, and the dark red region corresponds to the extrapolation in the mean-field unstable region.

\section{Numerical details}\label{app:numerics}
Ground states are calculated by solving Eq.~\eqref{eq:TeGPE} in imaginary time using the split-step Fourier method and an adaptive time step. We implement a grid consisting of $(N_x,N_y, N_z) = (128, 128, 64)$ grid points.
The dipolar interactions [second term of Eq.~\eqref{eq:Uint}] are efficiently evaluated in Fourier space thanks to the convolution theorem. 
In order to avoid artificial self-interaction of alias Fourier copies of the system, we ensure that the wave function has decayed to zero within one quarter of each of the grid edges.
The initial states of the imaginary time simulations, for the random initialization as well as for the first state of the hysteresis simulations, are initialized with a three-dimensional Gaussian scaled by a multiple of the harmonic oscillator length, perturbed by random Gaussian noise. The convergence of the states is ensured by the normalized residual, $r \equiv \int \mathrm{d}\mathbf{x}|H_\mathrm{TeGPE}\Psi - \mu\Psi|/\int \mathrm{d}\mathbf{x}\;\mu\Psi \leq 10^{-5}$, with $\mu$ the chemical potential.

\section{Density contrast for a trapped system}\label{app:contrast}
To calculate the contrast of a state, we first identify density peaks with at least $30\%$ of the maximum density and then use Delaunay triangulation to extract all edges to the nearest neighbors of density peaks. 
The density is then linearly interpolated along those edges, and the corresponding contrast $C$ for one edge is defined as 
\begin{align}
    C = \frac{n_\mathrm{max} - n_\mathrm{min}}{n_\mathrm{max} + n_\mathrm{min}},
\end{align}
where $n_\mathrm{max}$ is the average density of the two peaks enclosing the edge and $n_\mathrm{min}$ denotes the minimum density along the edge.
The mean of the contrasts along all edges is then taken as the contrast for the state. 
Since some states can exhibit density modulations in $x$ direction along a stripe, we only consider edges that are tilted by more than $30^\circ$ relative to the $x$ axis.

\section{Triangular-to-stripe order parameter}\label{app:classification}
\begin{figure}[t!]
    \centering
    \includegraphics[width=0.8\linewidth]{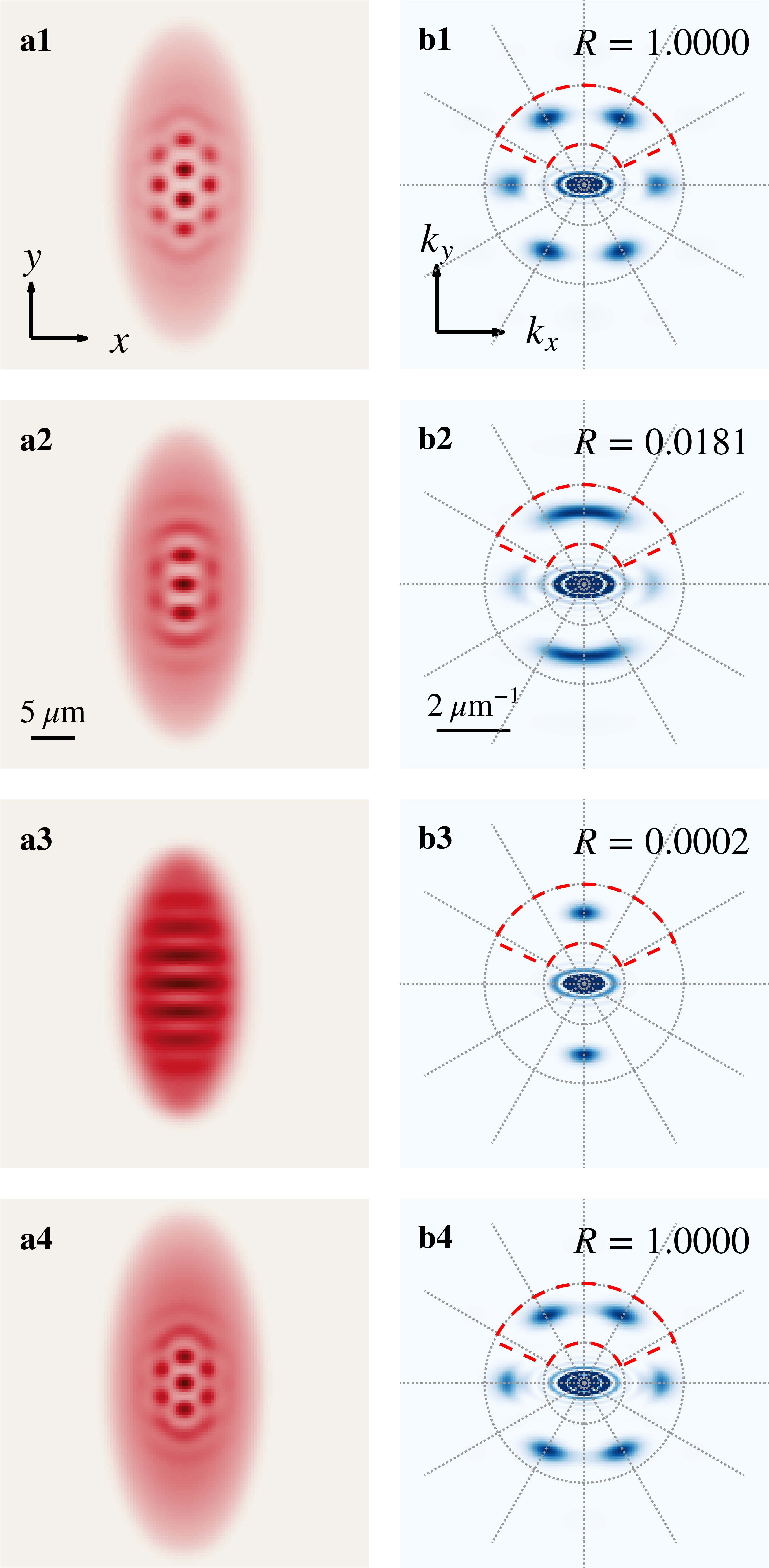}
    \caption{\textbf{Particle densities and power spectra.} Real-space densities (a1-a4) and corresponding power spectra (b1-b4) in the region of the unmodulated-to-modulated transition. States for $T = \SI{100}{\nano\K}$, $N = 2\times 10^5$ and $\nu_z = \SI{150}{\Hz}$ at $\alpha = 20$, $30$ and $50^\circ$ are shown in (a1-a3) and (b1-b3), respectively. A state for $T = \SI{60}{\nano\K}$, $N = 4\times 10^5$ and $\nu_z = \SI{150}{\Hz}$ at $\alpha = 15^\circ$ is shown in (a4, b4). Given are the $R$ values used for the distinction between triangular and stripe states.}
    \label{fig:PowerSpectra}
\end{figure}

To investigate the structural arrangements, we classify triangular and stripe states with an order parameter obtained from the two-dimensional power spectrum, 
\begin{align}\label{eq:PowerSpectrum}
    S(\mathbf{k}) = S(k_x, k_y) = S(k_\rho, \theta) = \left|\mathcal{F}\left[n(x, y, z = 0)\right]\right|^2, 
\end{align}
where $\mathcal{F}$ is  the two-dimensional Fourier transform and $\mathbf{k}$ the in-plane wave vector of Cartesian (polar) coordinates $k_x$ and $k_y$ ($k_\rho$ and $\theta$). We zero-pad the real-space densities to increase the momentum-space sampling. Fig.~\ref{fig:PowerSpectra} shows some exemplary real-space densities (a1-a4) and the corresponding power spectra (b1-b4) for different parameters in the region of the transition. 

The $R$ value used for the classification of triangular and stripe states is given by 
\begin{align*}
    R = \frac{S_2}{S_1},
\end{align*}
where $S_1$ and $S_2$ are the amplitudes of the two dominant peaks in the region $k_\rho \in [1.1, 2.7]\,\um^{-1}$ and $\theta \in [35^\circ,155^\circ]$, indicated by the red-dashed lines in Fig.~\ref{fig:PowerSpectra}. 
We further impose minimal radial and angular separations for $S_1$ and $S_2$ in $k_\rho$ and $\theta$ by $0.6\,\um^{-1}$ and $6^\circ$, respectively. See also~\cite{Chandrashekara2026Competing} for further details and experimental examples.

Triangular arrangements give an $R$ value close to one~[(a1, b1) and (a4, b4)] with peaks evenly spaced by $60^\circ$. Stripe states~(a3, b3) give $R$ values close to zero with peaks at $k_x = 0$, spaced by $180^\circ$. The state~(a2, b2) shows that a low $R$ value can also correspond to a geometry that appears more like a one-dimensional droplet chain with a diffuse halo. The density modulation in the outer halo is too weak to be considered as a crystal, and the state is identified as a stripe, i.e.~a one-dimensional modulated state.
We choose the threshold for the stripe classification to be $R<0.05$. We note that similar appearing real-space densities as in~(a2) and~(a4) can lead to distinct power spectra and classifications. This constitutes the ambiguity of the classification discussed in the main text.
We further note that the triangular state in~(a4) with $N=4\times 10^5$ and $\alpha = 15^\circ$ corresponds to a continuous unmodulated-to-modulated transition [Fig.~\ref{fig:hysteresis}(b)].

\section{Pancake trap}\label{app:pancake_trap}
\begin{figure}[ht!]
    \centering
    \includegraphics[width=1\linewidth]{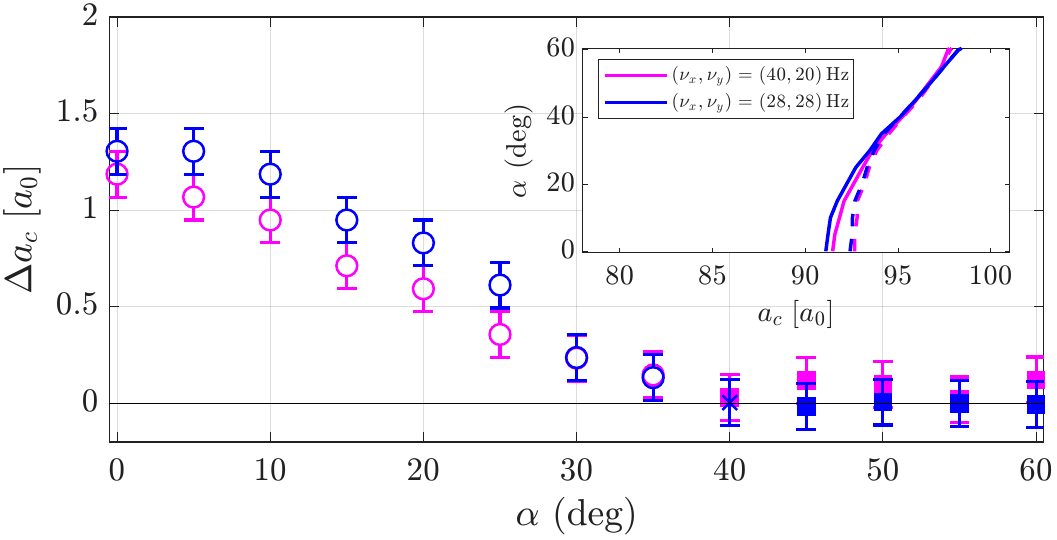}
    \caption{\textbf{Hysteresis and phase boundaries in a pancake trap.} Bistability range $\Delta a_c$ in a pancake trap with $\nu_\perp = \SI{28}{\Hz}$ (blue) compared to the surfboard trap with $(\nu_x, \nu_y) = (40, 20)\, \mathrm{Hz}$ (magenta) with the reference parameters $N = 2\times 10^5$, $T = \SI{60}{\nano \K}$ and $\nu_z = \SI{150}{\Hz}$ for both traps. The insets show the critical scattering lengths for the ingoing (solid) and outgoing (dashed) directions.}
    \label{fig:hysteresis_pancake}
\end{figure}

In order to investigate the influence of the trap isotropy, we change to a pancake trap with $\nu_x = \nu_y = \nu_\perp$. To maintain the position of the unmodulated-to-modulated phase boundaries from the surfboard trap, we follow the results from Ref.~\cite{Poli2021Maintaining} and choose $\nu_\perp = \sqrt{\nu_x \nu_y} \approx \SI{28}{\Hz}$ and perform the same hysteresis analysis for the reference parameters $N = 2\times 10^5$, $T = \SI{60}{\nano \K}$ and $\nu_z = \SI{150}{\Hz}$. Fig.~\ref{fig:hysteresis_pancake} shows the bistability range $\Delta a_c$ and phase boundaries (insets) for the surfboard (magenta) and pancake (blue) traps. The isotropy of the trap neither changes significantly the hysteresis nor the structural arrangement of the states. Related to our choice of $\nu_\perp$, the phase boundaries only slightly moved in scattering length.

\bibliography{References}

\end{document}